\pdfoutput=1 
\documentstyle[12pt,graphicx,natbib,hyperref]{article}
\def\sun{\hbox{$\odot$}}
\def\aap{A\&A\,  }
\def\aaps{A\&AS  }
\def\aj{AJ  }
\def\apj{ApJ\,  }
\def\apjs{ApJS  }
\def\cjaa{Chinese J. Astron. Astrophys.  }

\def\mnras{MNRAS\,  }
\def\pasa{PASA  }

\def\pla{Phys. Lett. A   }
\def\prd{Phys. Rev. D   }
\def\pre{Phys. Rev. E   }
\begin{document}
\title
{
Revisiting the Cosmological Principle
in  a Cellular  Framework
}
\author
{
L. Zaninetti             \\
Dipartimento di Fisica   , \\
           Via Pietro Giuria 1 ,  \\
           10125 Torino, Italy   
}

\maketitle
\section*{}

The Cosmological Principle  in it's various versions
states that: (i) the Galaxy does not occupy a particular
position, (ii) the Universe is homogeneous and isotropic. 
This statement does not agree  with the recent astronomical 
observations  in the range   z lower than 0.05  which are
in agreement  with a cellular structure  of the  Universe. 
Here we present  a local analysis of the 
inhomogeneity of  the
Universe.  
When  z is greater than  0.05  our analysis cannot be applied
because the astronomical sample of galaxies here 
processed  is not 
complete.  
The two tools of the Poisson Voronoi tessellation (PVT) 
and  the luminosity
function for galaxies allow 
of building a new version of the
local Cosmological Principle.
\\
keywords                    \\
         {
Cosmology: miscellaneous 
Cosmology: observations 
Cosmology: theory 
}
\section{Introduction}

All the various theories which belong
to cosmology are based on the
``Cosmological Principle''
which is now presented in two versions.
The first version asserts  that
``There is nothing special about our  location
in the Universe,''
see  \cite  {Ryden2003,Oxford_Cosmology_2009}.
The second version  is
 ``Viewed on a sufficiently large scale,
the properties of the Universe are the same for all observers,''
see \cite{Keel2007}. Other  versions
 which allow the formation of
structures and incorporate probability 
distributions can be found
in \cite{Schwarz2009}.

These two statements are strictly connected
with the concept of isotropy and 3D homogeneity
of the Universe.
In the following we will show
that
\begin{itemize}
\item  The Local Super Cluster (LSC), the nearest
       part of the universe,  has a dimension
       around 2 which is characteristic of a bi-dimensional
       space  and therefore  the concept
       of 3D homogeneity is violated by the observations.
\item  The concept  of isotropy 
       in  the spatial distribution
       of galaxies  is not confirmed due to the
       3D cellular structure  of galaxies.
\end{itemize}

In order to demonstrate  the previous two statements
this paper is organized as follows.
Section \ref{elementary} reviews
the adopted cosmology.
Section \ref{voronoi} reports a spherical cut
of the 3D Voronoi diagram
as  well an introduction 
to the thick Voronoi faces.
Section \ref{catalogue}
reports a spherical cut
of an astronomical catalog.
Section \ref{luminosity} reports the most used
luminosity function
for galaxies,
a modern version of
the Malmquist bias, and the
measure of the spatial dimensions of the LSC.
Section \ref{conclusions} reports
a revised version of the
``Cosmological Principle.''

\section{Elementary Cosmology}
\label{elementary}

Starting from
\cite{Hubble1929}
the suggested correlation
between expansion velocity  and distance is
\begin {equation}
V= H_0 D  = c \, z
\quad,
\label {clz}
\end{equation}
where the Hubble constant  is
$H_0 = 100 h \mathrm{\ km\ s}^{-1}\mathrm{\ Mpc}^{-1}$,
with $h=1$
when  $h$ is not specified,
$D$ is the distance in Mpc,
$c$ is  the  velocity  of light, and
$z$   is  the redshift.
The redshift of galaxies
is explained by  the Doppler effect
or by  alternative theories  such as
the presence of a
hot plasma with low density, see \cite{Brynjolfsson2009}.
In a Euclidean, non-relativistic,
and homogeneous universe,
the flux of radiation,
$ f$,  expressed in $ \frac {L_{\sun}}{Mpc^2}$ units,
where $L_{\sun}$ represents the luminosity of the sun,
is
\begin{equation}
f  = \frac{L}{4 \pi D^2}
\quad,
\label{flux}
\end{equation}
where $D$   is  the distance of the galaxy
expressed in Mpc,
and
\begin{equation}
D=\frac{c_l z}{H_0}
\quad  .
\end{equation}

\section{Voronoi Diagrams}

\label{voronoi}

The points of a tessellation in 3D are of four types,
depending
on how many nearest
neighbors in $ES$, the ensemble of seeds, they have.
The name seeds  derive  from their  role in generating cells.
Basically  we have two kinds  of seeds , Poissonian and non
Poissonian  which generate the Poissonian Voronoi tessellation
(PVT) and the non Poissonian Voronoi tessellation (NPVT). The
Poissonian   seeds  are generated independently on the $X$, $Y$
and $Z$ axis in 3D through a subroutine  which returns a
pseudo-random real number taken from a uniform distribution
between 0 and 1 . This  is the case most studied and for 
practical
purposes,
 the subroutine
RAN2  was used, see \cite{press}.
The non Poissonian   seeds  can  be  generated
in an infinite  number  of different ways :
some examples of  NPVT  are reported  in  \cite{Zaninetti2009c}.

A point with exactly one nearest neighbor's is in
the interior of a cell,
a point with two nearest neighbors is on the face between
two cells,
a point with
three nearest neighbors is
on an edge shared by three cells,
and a point with four neighbors is a
vertex where three cells meet.
The Abell clusters and therefore the galaxies were  originally
inserted  on the Voronoi vertexes, which are the points which
share the same distance  from four seeds, see \cite{Weygaert1989}.
The target to reproduce  in the previous analysis was  the
cluster-cluster  correlation function. In order to explain the
cellular  structure of the universe the galaxies were later on
collocated on the faces of a  NPVT in order  to reproduce the
galaxy-galaxy correlation function and the cluster-cluster
correlation function,
 see \cite{zaninettig}.
More detailed  simulations of the various  astronomical
catalogs  as well of the
galaxy-galaxy correlation function
were  recently  obtained inserting the
galaxies on the faces of a PVT,
see \cite{Zaninetti2006,Zaninetti2010a}.

According to this trend we will assume
that the galaxies are situated on the faces of
the PVT.
Some of the properties of  the  PVT may  be deduced
from approximate  arguments
introducing  the averaged radius of a polyhedron ,
$\bar {R}$  and the averaged diameter
$\bar {D}=2\bar {R}$.
This theoretical quantity has   its observational
counterpart in the averaged  diameter of the
voids between galaxies,
$\overline{D^{obs}}$.
A careful analysis  of
the effective  radius of the voids 
between galaxies  in 
SDSS DR7, see Section \ref{sec_statistics},  
derives   $\overline{D^{obs}}=\frac{36.46}{h}~Mpc $.
The averaged volume is
\begin{equation}
\bar {V} = \frac{4}{3} \pi \bar {R}^3
\quad,
\end{equation}
and the averaged number of faces is
$n$ which according to \cite{okabe} is $n$ = 15.35~.

\begin{figure}
\begin{center}
\includegraphics[width=6cm]{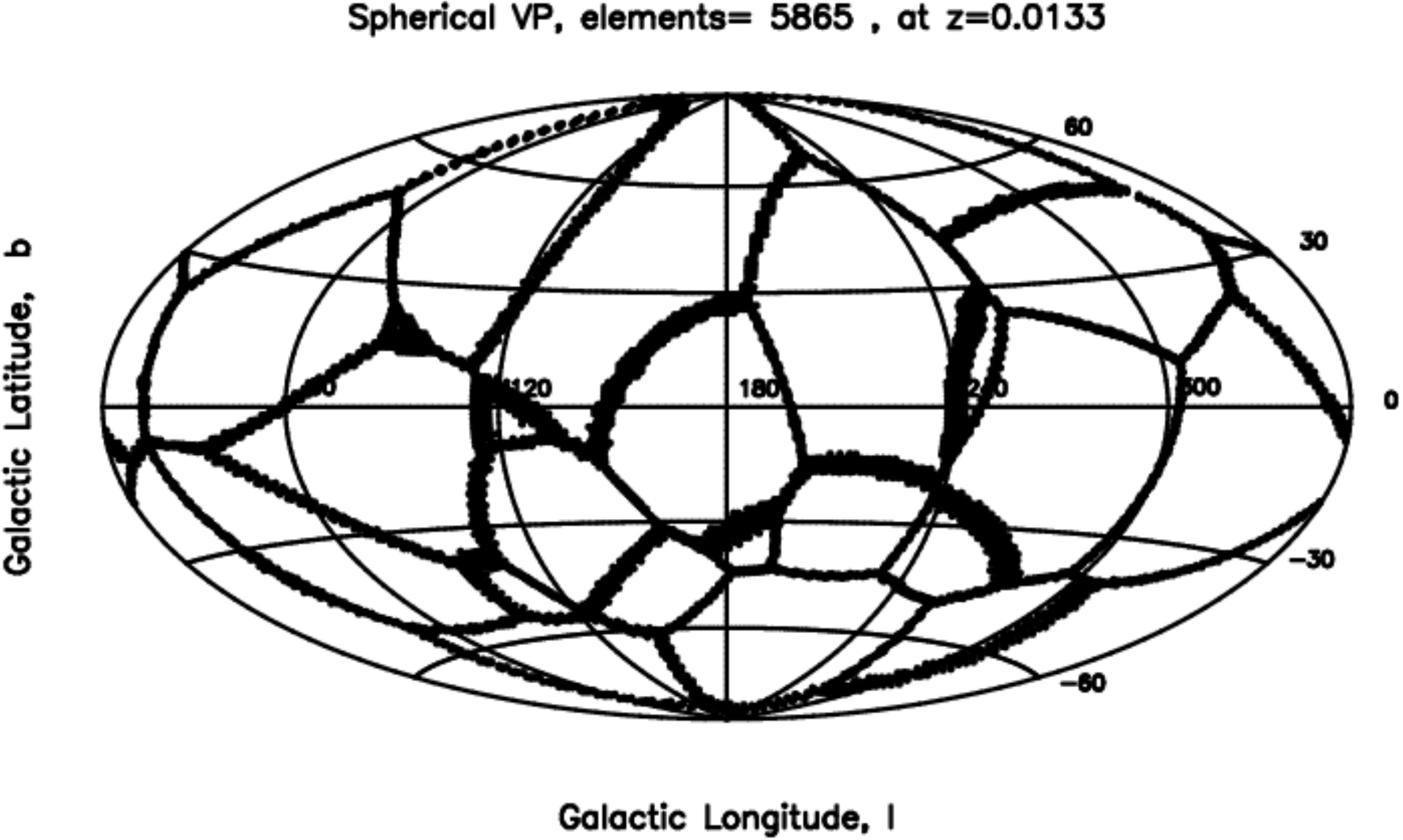}
\end {center}
\caption
{
The Voronoi diagram $V_s(2,3)$
in the Hammer--Aitoff  projection
$0.0133  \leq z \leq 0.0144$
or
56.9 Mpc  $\leq D  \leq 61.67$ Mpc
when  $\bar{R}$=18.16 Mpc.
}
          \label{aitof_sphere}%
    \end{figure}

\begin{figure}
\begin{center}
\includegraphics[width=6cm]{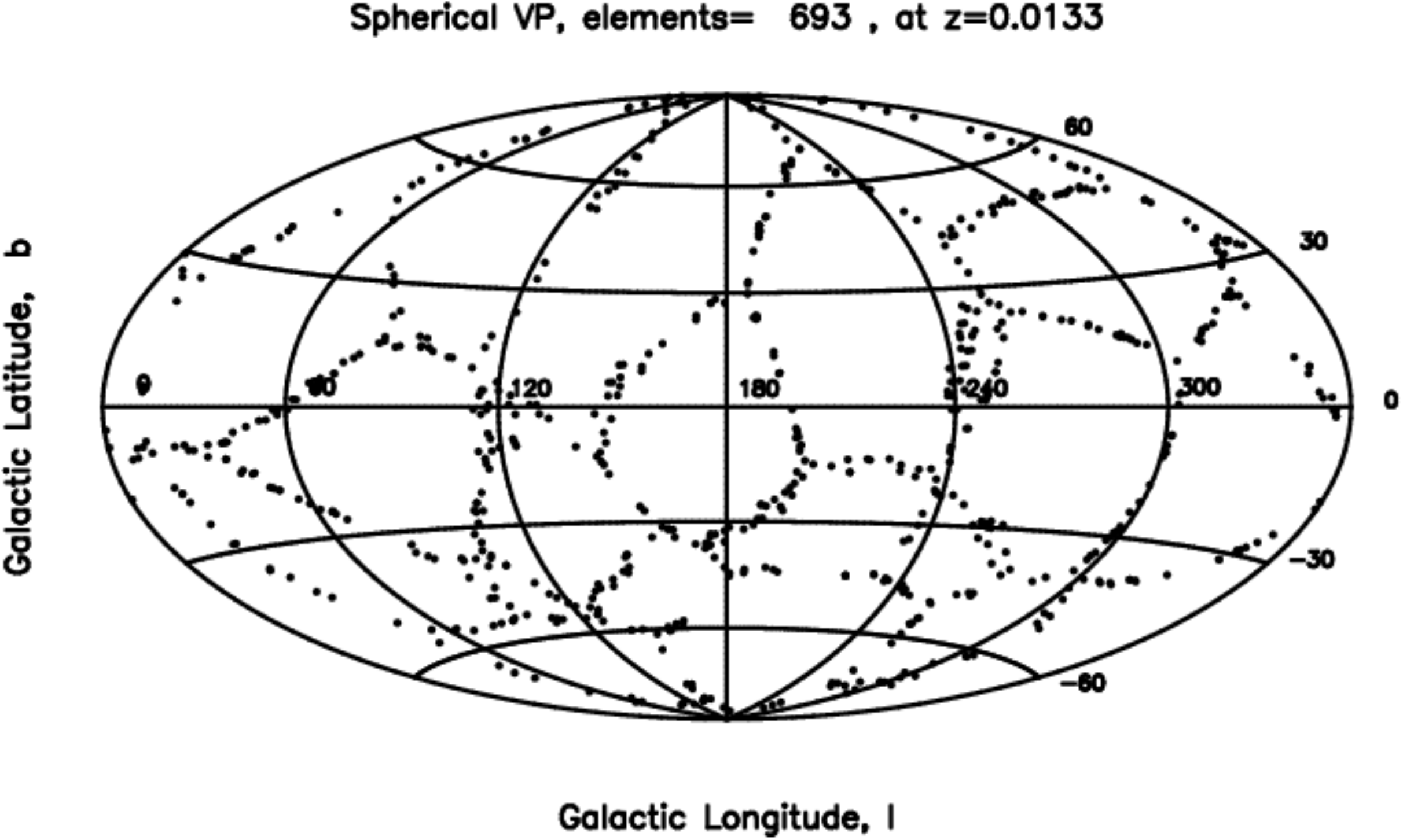}
\end {center}
\caption
{
The Voronoi diagram $V_s(2,3)$
in the Hammer--Aitoff  projection
when 
693 galaxies are extracted  from the network
of  Figure \ref{aitof_sphere}.
}
          \label{aitof_sphere_sel}%
    \end{figure}

The averaged  surface 
area of a
polyhedron, $\bar {S} $, is
\begin{equation}
\bar {S} = 4 \pi \bar{R}^2
\quad,
\end{equation}
and the approximate area of a face,
$\bar {A} $,
is
\begin{equation}
\bar {A} = \frac{ 4 \pi \bar{R}^2 } {n}
\quad .
\end{equation}
The approximate side of a face, $ \bar {l} $, is
\begin{equation}
\bar {l}
=
\sqrt {
\frac{ 4 \pi \bar{R}^2 } {n}
}
\quad .
\end{equation}
On assuming  that our galaxy is at the center
of an irregular face of a PVT,
the galaxies
on the other faces are comprised 
within a  distance $D_F$
\begin{equation}
D_F \approx  \frac {\bar {l}} {2}
\quad .
\label{firstface}
\end{equation}
The cross sectional area of a  PVT can
also be visualized through
a spherical cut characterized by a constant value
of the distance from  the center of the box,
in this case expressed in $z$ units,
see Figure~\ref{aitof_sphere};
this intersection is called $V_s(2,3)$  where the
index $s$ stands for sphere.
The galaxies are thought to be situated
on the faces of a PVT network  and
Figure~\ref{aitof_sphere_sel} reports
an example of a   spherical cut.

An astronomical  slice consists of a pie diagram with polar angle
given by the  right  ascension and  polar  radius  given  by the
redshift. An  example  of a slice of the Sloan Digital Sky Survey
(SDSS)  which  contains all
the galaxies within an  opening 
angle in declination ,$\varphi$ ,
of $2.5^{\circ}$ around  declination =0
can be found at
$http://www.sdss3.org/gallery/sdss\_pie2.php$
.
 Other surveys have  the
following opening angle in declination : $3^{\circ}$ for   the
second Center for Astrophysics (CFA2) survey,
 see  \cite{geller} ,
$4^{\circ}$ for   the 2dF Galaxy Redshift Survey (2dFGRS),
 see
\cite{Colless2001} and  $4^{\circ}$ 
for the Six-degree Field
instrument (6dF),
  see \cite{Jones2004}. Particular attention
should be paid to the fact that the astronomical slices are not a
plane which intersects a Voronoi Network.

\begin{figure}
\begin{center}
\includegraphics[width=6cm]{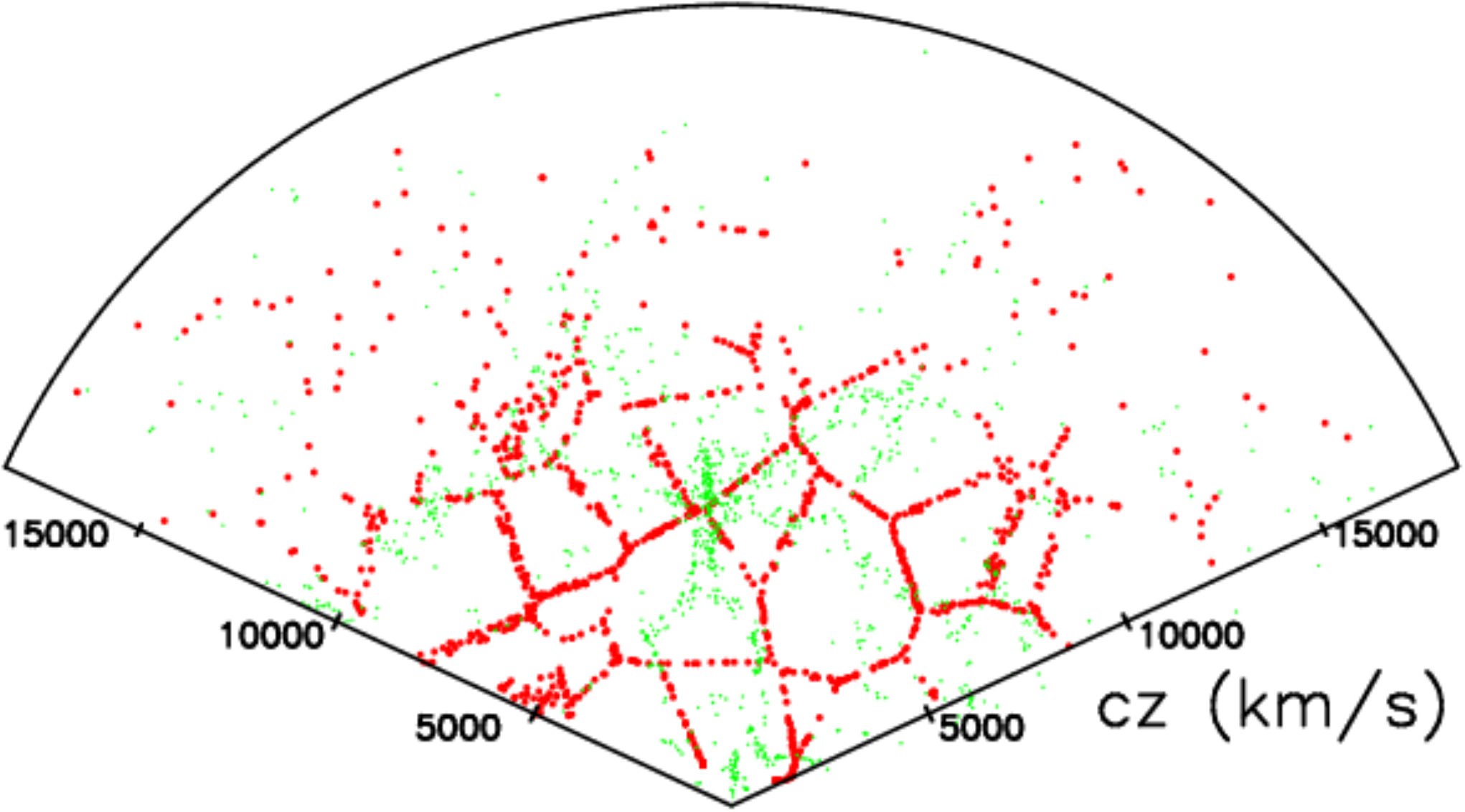}
\end {center}
\caption
{
Polar plot of   the   real galaxies (green  points)
belonging to the second CFA2 redshift catalog
and  the simulated galaxies in the  PVT 
framework (red points).
More details can be found  in 
\cite{Zaninetti2006}.
}
          \label{true_simu_color}%
    \end{figure}

In order to quantify this effect we introduce a confusion
 distance, $D_c$, as the distance  after which 
the half altitude of the slices equalizes the 
observed average diameter  $\overline{D^{obs}}$
\begin{equation}
D_c \tan (\frac{\varphi}{2}) = \frac{1}{2} \overline{D^{obs}}
\quad ,
\end{equation}
where $\varphi$ is the opening angle  of the slice 
and  $\overline{D^{obs}}$ the averaged diameter of
the voids.
In the case of  SDSS   $\varphi=2.5^{\circ}$ 
and therefore 
$D_c=835.4/h ~Mpc$
when    $\overline{D^{obs}}=\frac{36.46}{h}$~Mpc.
For distances greater than $D_c$ 
the voids in the distribution
of galaxies are dominated by the confusion.
For distances lower  than $D_c$
the filaments of galaxies  can be considered  
the intersection
between a plane and the faces of the Voronoi Polyhedrons.
The  PVT  offers a new classification 
for extragalactic aggregates.
\begin{enumerate}
\item
Groups and  clusters of galaxies 
are  situated  on the faces  of the PVT 
with typical linear dimensions of  $\approx$ 16 Mpc.
\item
The local  super-cluster is made by the network  
of many faces of the PVT.
For distances  equal  to or slightly greater than
$\overline{D^{obs}}/2$, a spherical  cut  should reveal
$\approx$ 16  voids .
\item
The filaments  of galaxies visible 
in the slices of oriented catalogs  are due  
to the intersection between a plane  and the PVT
network of faces as first approximation.
An improvement can be obtained   by coding
the intersection between the slice of a given  opening 
angle  and the PVT network of faces, 
see \cite{Zaninetti2006,Zaninetti2010a}. 
As an example
Figure \ref{true_simu_color}  
reports  both  the 
CFA2 slice as  well the simulated slice.
\item
The Great Wall, see  \cite{geller}, is due to a combined
effect  of the  photometric maximum,
see  Equation 
(\ref{posmaximum})  
and the PVT network
of faces.
\end{enumerate}

The cellular structure in the spatial
distribution of galaxies
can be generated
in other approaches such as
the long-range force acting only 
between non-baryonic particles,
see \cite{Peebles2010}.

\subsection{Thickness}

\begin{figure}
\begin{center}
\includegraphics[width=6cm]{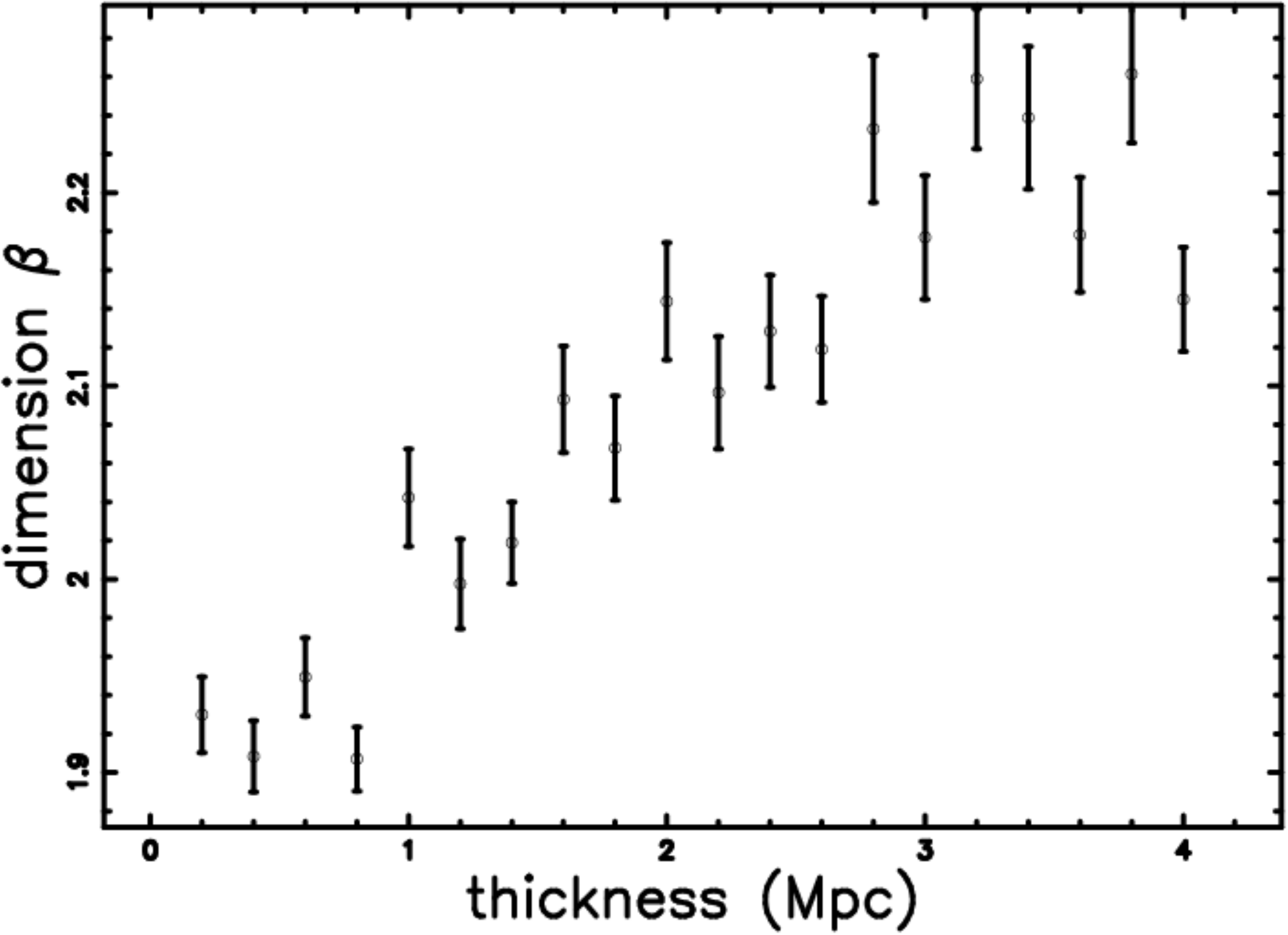}
\end {center}
\caption
{
The  dimension of the number of  galaxies
represented with vertical error bars
as  function of the thickness   
in Mpc.
}
          \label{areathick}%
    \end{figure}

The concept  of thickness of a face can be  introduced 
by a numerical  experiment.
We generate $N_r$ random galaxies 
in 2D and then we progressively 
shift toward  3D increasing the thickness of the 
box from zero to a given value.
The analysis of the number of galaxies,
 $N(R)$,  as  function of the distance 
can be done by assuming  a  power law
dependence  of the type
\begin{equation}
N(R) = C R^{\beta}
\label{rpower}
\quad .
\end{equation}
The two parameters $C$ and  $\beta$ can be found
from the following
logarithmic transformation
\begin{equation}
\ln(N(R)) = \ln(C) + \beta \ln (R)
\quad,
\end{equation}
which can be written as
\begin{equation}
y  = a_{LS} +b_{LS}x
\quad  .
\end{equation}
The application
of the least squares method
through the FORTRAN subroutine LFIT from
\cite{press} allows of finding $a_{LS}$, $b_{LS}$ and the
errors $\sigma_a$ and $\sigma_b$.
The resulting dimension   
increases  progressively  with thickness,
see Figure \ref{areathick}.

\begin{figure*}
\begin{center}
\includegraphics[width=6cm]{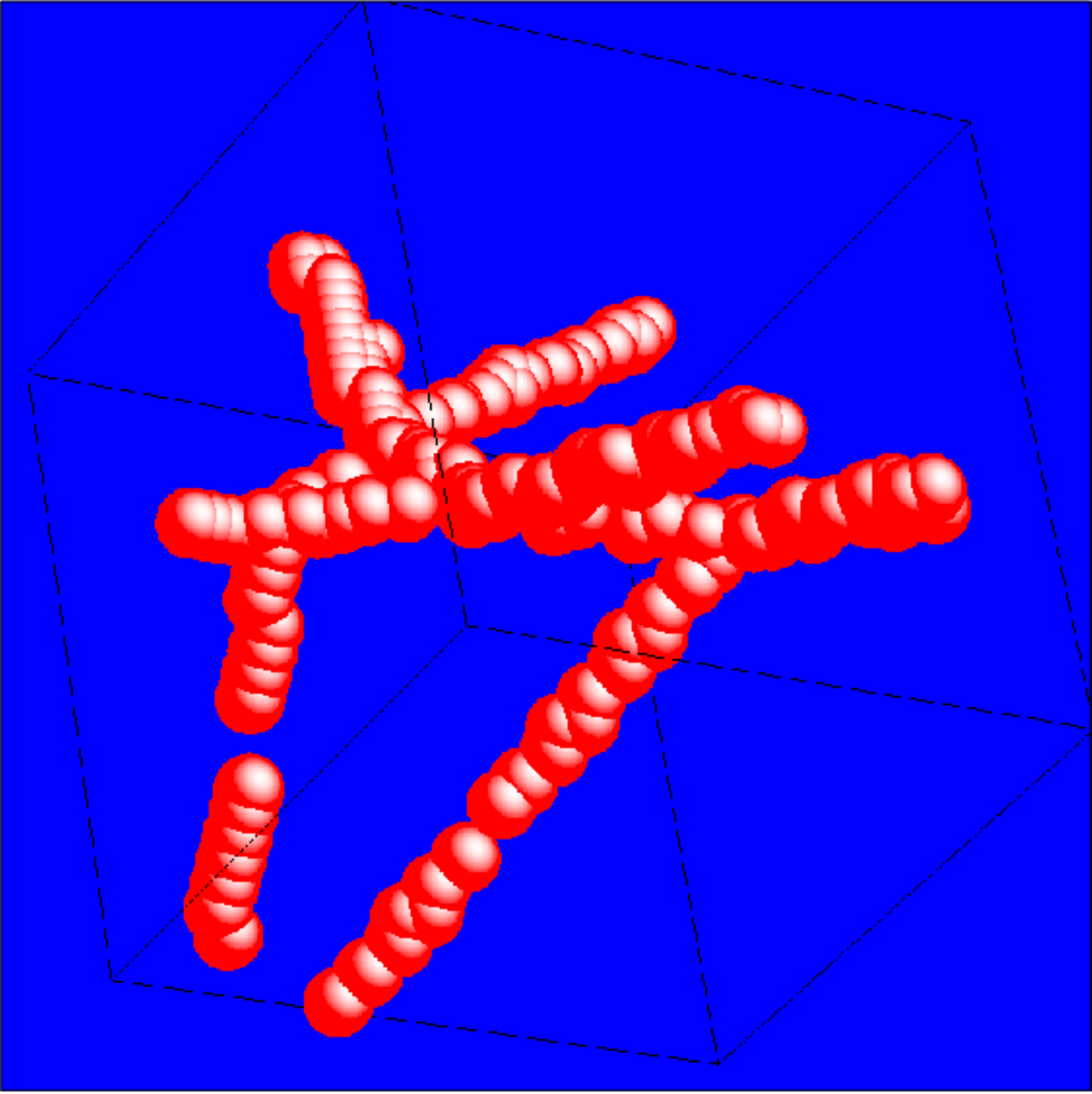}
\end {center}
\caption{
3D visualization of the 
edges of the Poissonian Voronoi--diagram.
The  parameters
are      $ pixels$= 60, $ N_s   $   = 60, 
         $ side  $   = 89 $Mpc$, 
         $h=0.7 $  and    $ amplify$= 1.2.}
          \label{spigoli3d_sb}%
    \end{figure*}

\begin{figure*}
\begin{center}
\includegraphics[width=6cm]{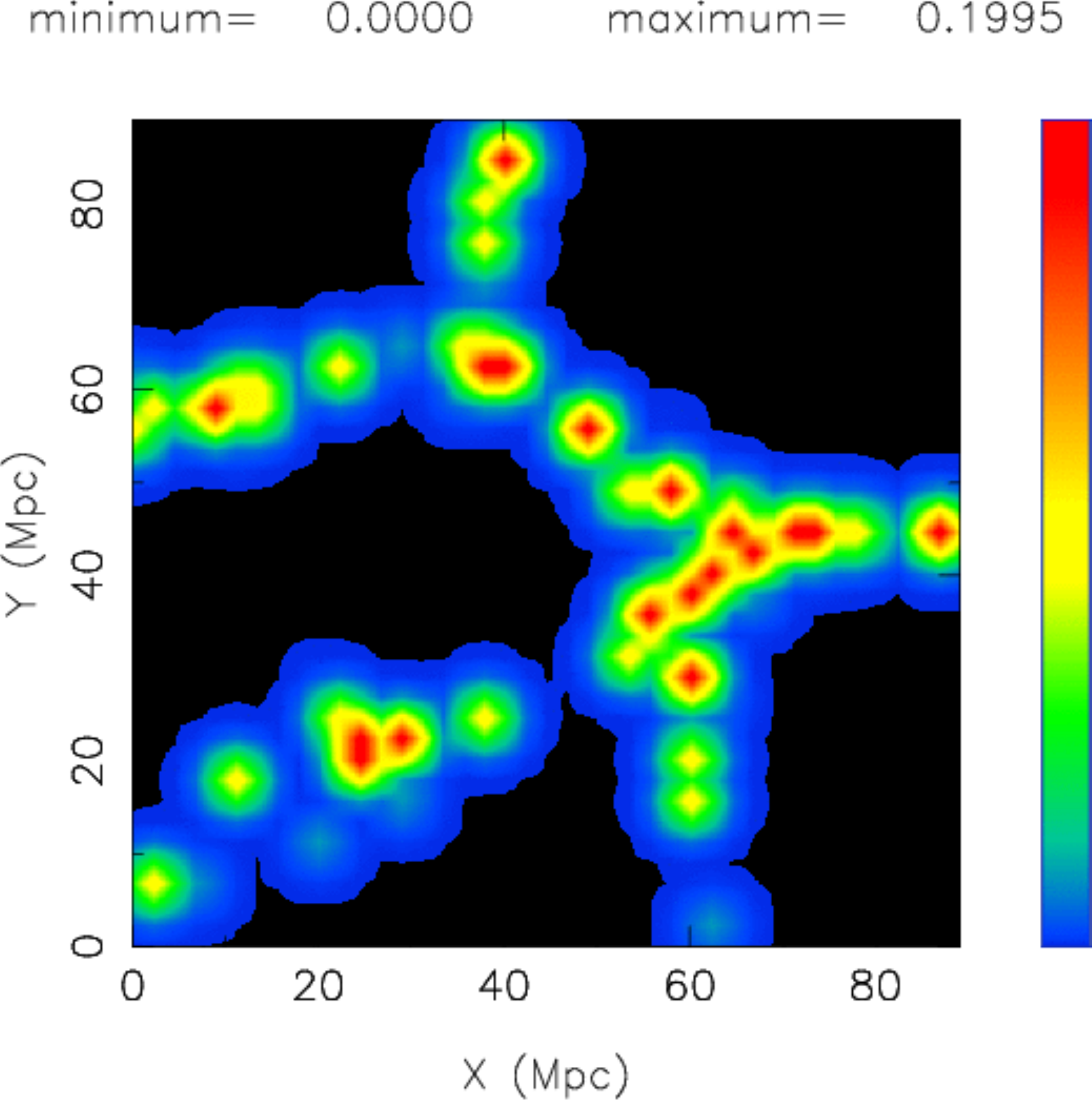}
\end {center}
\caption{
Cut  in the middle of the 3D grid $P(i,j,k)$
which represents a theoretical 2D map 
of the probability of having  
a galaxy. 
The  Voronoi parameters are the same as in  
Figure~\ref{spigoli3d_sb} and $\sigma=0.8 Mpc$.
The X and Y units are in Mpc.
        }
    \label{probability2d}
    \end{figure*}

The fact that the  dimension is not exactly 
2 at thickness =0 ,
more exactly  we have  $\beta=1.8$ when 
$N_r$=10000 , is due to the cardinality
of $N_r$ , as an example we have $\beta=1.96$ when 
$N_r$=1000000.
This first analysis  indicates  that the dimension 
is function both of a geometrical factor  as thickness
and numerical factor , i.e. the number of considered 
galaxies.
We now outline the method that allow us 
to  insert the galaxies on the   thick faces, 
see \cite{zaninetti95,Zaninetti2010a}.
A practical implementation is to model  a decreasing probability
of  having  a galaxy  in the direction  perpendicular to the face.
As an example we assume  a  probability, 
$p(x)$, of  having  a galaxy 
outside the face distributed as a Normal (Gaussian) 
distribution
\begin{equation}
p(x) = 
\frac {1} {\sigma (2 \pi)^{1/2}}  \exp {- {\frac {x^2}{2\sigma^2}}} 
\quad  ,
\label{gaussian}
\end{equation}
where $x$ is the distance in $Mpc$ from the face and $\sigma$ 
the standard deviation in $Mpc$.
Once the complex 3D behavior of the faces of the Voronoi
Polyhedron is set up  we can memorize 
such a probability on a 3D grid $P(i,j,k)$ 
which  can be found in the following way  
\begin{itemize}
\item In each lattice point $(i,j,k)$ we search for  
      the nearest  element
      belonging to a Voronoi face. The probability of having  
      a galaxy
      is therefore computed according to formula~(\ref{gaussian}).
\item    A number of galaxies, $N_G=n_* \times side^3$
         is then inserted in the box; 
         here $n_*$ represents the   density of  galaxies 
\end{itemize}
Figure~\ref{spigoli3d_sb} visualizes  
 the edges belonging  to the Voronoi 
diagrams and Figure~\ref{probability2d} represents 
a cut in the middle of the probability, $P(i,j,k)$, 
of having  a galaxy to a given 
distance from a face.

We are now ready  to evaluate  the spatial dimension
in two  cases   
\begin{itemize}
\item ntrials  centers   
      randomly situated   on a face ,
\item ntrials  centers   
      randomly situated   on a edge,
\end{itemize}
see Table  \ref{tabledimension}.
The case  of evaluation starting from 
the edges produces  a bigger dimension because we 
enclose three faces  rather than one.

\begin{table}
 \caption[]
{
The dimension in two cases
when ntrial=10.
}
 \label{tabledimension}
 \[
 \begin{array}{lc}
 \hline
 \hline
 \noalign{\smallskip}
case             &   value                          \\ \noalign{\smallskip}
faces               &  1.65 \pm 0.1 \\ \noalign{\smallskip}
edges               &  1.79 \pm 0.07 \\ \noalign{\smallskip}
 \hline
 \end{array}
 \]
 \end {table}

\section{Observations}

This section processes  the 
Sloan Digital Sky Survey Data Release 7 (SDSS DR7)
and  the 
Two-Micron All Sky Survey (2MASS).  
\subsection{Observed Statistics of the Voids}
\label{sec_statistics}
The distribution  of the effective radius  between the galaxies
of  SDSS DR7
has been reported  in 
\cite{Vogeley2012} 
and  is also possible  to  extract  the catalog
which is visible
at \begin{verbatim}
http://www.physics.drexel.edu/~pan/voidcatalog.html
\end{verbatim}
; this catalog  contains   1054  voids  
and Table \ref{statvoids} reports their basic
statistical  parameters. 
\begin{table}
 \caption[]
{
The statistical  parameters 
of the effective radius in  SDSS DR7.
}
 \label{statvoids}
 \[
 \begin{array}{lc}
 \hline
 \hline
 \noalign{\smallskip}
parameter            &   value                          \\ \noalign{\smallskip}
mean                       &  18.23h^{-1}~ Mpc   \\ \noalign{\smallskip}
variance                   &  23.32h^{-2}~ Mpc^2 \\ \noalign{\smallskip}
standard~ deviation         &  4.82h^{-1} ~ Mpc   \\ \noalign{\smallskip}
kurtosis                   &  0.038        \\ \noalign{\smallskip}
skewness                   &  0.51         \\ \noalign{\smallskip}
maximum ~value              &  34.12h^{-1}~ Mpc   \\ \noalign{\smallskip}
minimum ~value              &  9.9h^{-1}~   Mpc   \\ \noalign{\smallskip} \hline
 \hline
 \end{array}
 \]
 \end {table}

\subsection{The 2MASS}
\label{catalogue}
The 2MASS   is a catalog
of galaxies which has instruments in the near-infrared
J, H and K-bands
(1-2.2 $\mu$\ m)
 and therefore detects  the galaxies
in the so called ``Zone of Avoidance,''
see \cite{Jarrett2004,Huchra2007}.
At  the moment  of writing
the  2MASS Redshift Survey  (2MRS)
consists of  44599  galaxies
with redshift in the interval $0\leq z\leq 0.09$,
see  \cite{Huchra2012}.

\begin{figure}
\begin{center}
\includegraphics[width=6cm]{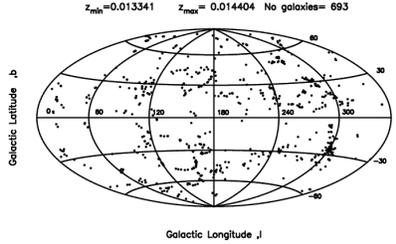}
\end {center}
\caption
{
Hammer--Aitoff  projection in galactic coordinates
of a spherical cut of the LSC  data 
at
$0.0133  \leq z \leq 0.0144$
or
$56.9 Mpc  \leq D  \leq 61.67$
.
}
          \label{2mrs_data}%
    \end{figure}

Figure \ref{2mrs_data} reports a spherical cut
at a given radius  in $z$
of the  LSC  according to
2MRS, which
is available  online  at   \\
https://www.cfa.harvard.edu/~huchra/seminar/lsc/lsc.dat\,.
The  LSC is centred  on the Virgo cluster , has a flattened structure
and  it's radius is  $\approx$ 40 Mpc.

A way to smooth  the results  of
Figure \ref{2mrs_data}   is through contour
of  isodensity of galaxies
in a cube of sides equal to 16 Mpc at a given
distance, see Figure~\ref{contour}.
\begin{figure}
\begin{center}
\includegraphics[width=6cm]{f08.pdf}
\end {center}
\caption
{
Logarithmic contours of density
of galaxies of the Large Super Cluster.
The density is evaluated in  cubes
of side = 16 Mpc
at
$z=0.0133 $
or
$D=56.9 Mpc$.
}
          \label{contour}%
    \end{figure}
The  previous  figure contains voids  in which  
the  density of galaxies  is zero.
In order to have in all  regions of the spherical cut  
a minimum density of  1 galaxy/(chosen~volume) we consider
the density as evaluated  in a cube which has 
a  side equal  the averaged diameter as given
by SDSS DR7, see  Figure \ref{rho}.
Now the density of galaxies has a minimum value of 1 
at  the center of a void and a maximum value of 270 
on a side  which  comes out  from the intersection 
of a sphere with a PVT.
This means that the density  of galaxies  varies  over
more than  two decades  on a spherical cut.

\begin{figure}
\begin{center}
\includegraphics[width=6cm]{f09.pdf}
\end {center}
\caption
{
Linear contours of density
of galaxies of the Large Super Cluster.
The density is evaluated in  cubes
of side = 16 Mpc
at
$z=0.0133 $
or
$D=56.9 Mpc$.
}
          \label{rho}%
    \end{figure}
\begin{figure}
\begin{center}
\includegraphics[width=6cm]{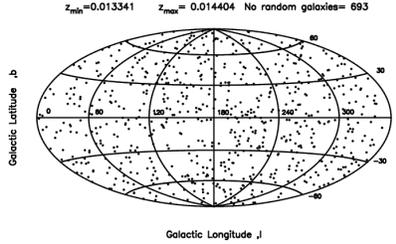}
\end {center}
\caption
{
Hammer--Aitoff  projection in galactic coordinates
of a spherical cut of a random sample  of galaxies  
at
$0.0133  \leq z \leq 0.0144$
or
$56.9 Mpc  \leq D  \leq 61.67$
.
The number of galaxies  is the same of Figure~\ref{2mrs_data}.
}
          \label{2mrs_data_random}%
    \end{figure}

The concept  of homogeneity  can be associated to a random sample
which has  the same cardinality  of the astronomical  galaxies but
placed at random in the same analyzed shell;
Figure \ref{2mrs_data_random}
reports an example of such random galaxies.
 The ratio
\begin{equation}
H(r)  = \frac{N_G(r)}{\langle N_R(r) \rangle} \label{formulah}
\quad ,
\end{equation}
quantifies  the homogeneity of the astronomical  sample with $
N_G(r)$  representing the number of real galaxies and  $N_R(r)$
the random galaxies in a cube of side $2r$ , $\langle N_R(r)
\rangle$ represents the  number of random galaxies averaged over
 $91\times181$ directions. 
A value of  of $H=1$ means  that the
sample is homogeneous  , a value  lower that 1 means  that the
sample is not homogeneous. 
An example  of the  various  value of
$H$ is  reported  in Figure  \ref{2mrs_data_random} and Figure
\ref{survival} reports  the numerical survival function for the
grid  points. 
From this last figure is possible to conclude that
we have non homogeneity  over the 68~$\%$ of the grid points of
the spherical  cut.

\begin{figure}
\begin{center}
\includegraphics[width=6cm]{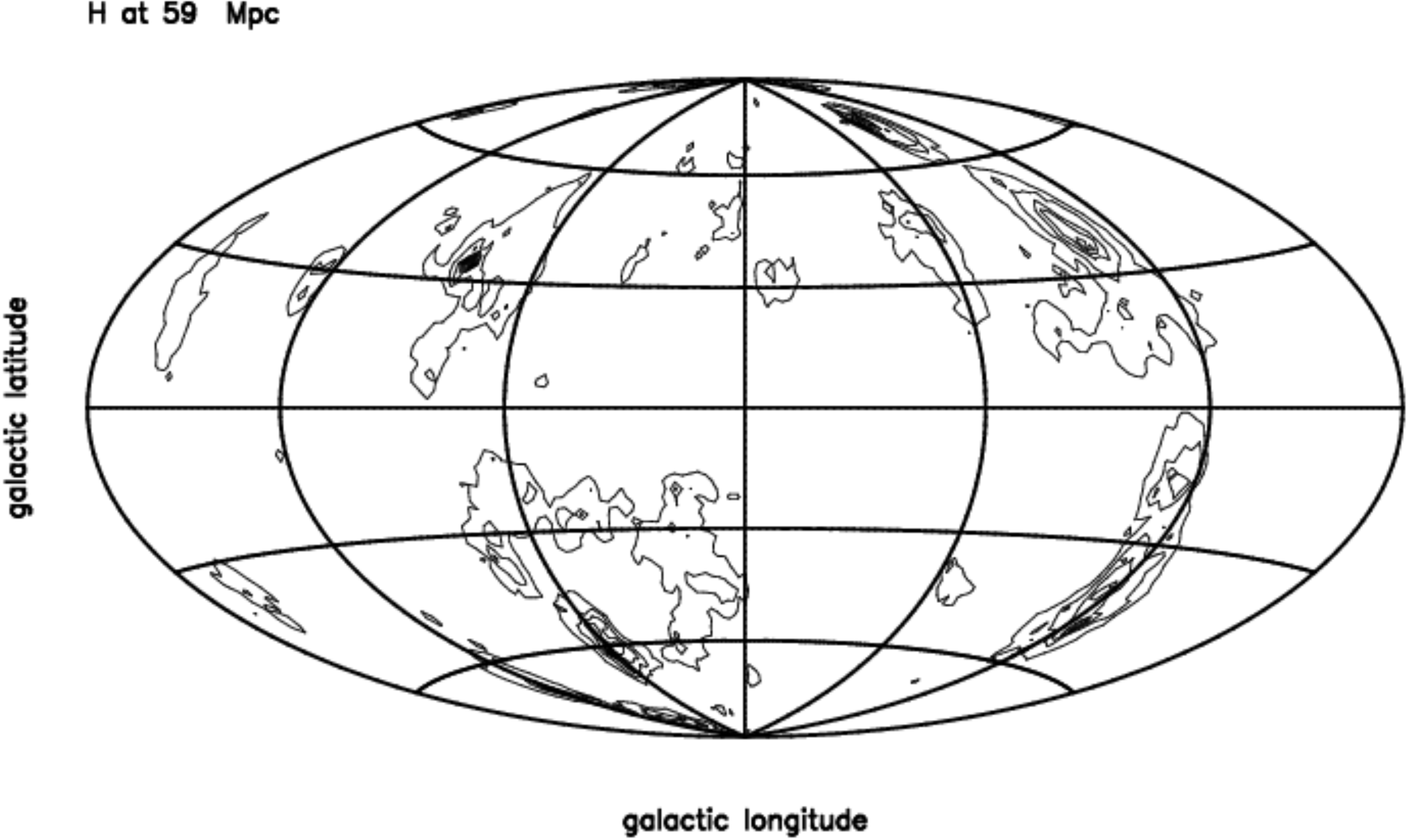}
\end {center}
\caption
{
Linear contours of  H as  given  by  (\ref{formulah}) 
in  the Large Super Cluster.
The density of the astronomical and random samples are
evaluated in  cubes
of side = 16 Mpc
at
$z=0.0133 $
or
$D=56.9 Mpc$.
}
          \label{comparison}%
    \end{figure}

\begin{figure}
\begin{center}
\includegraphics[width=6cm]{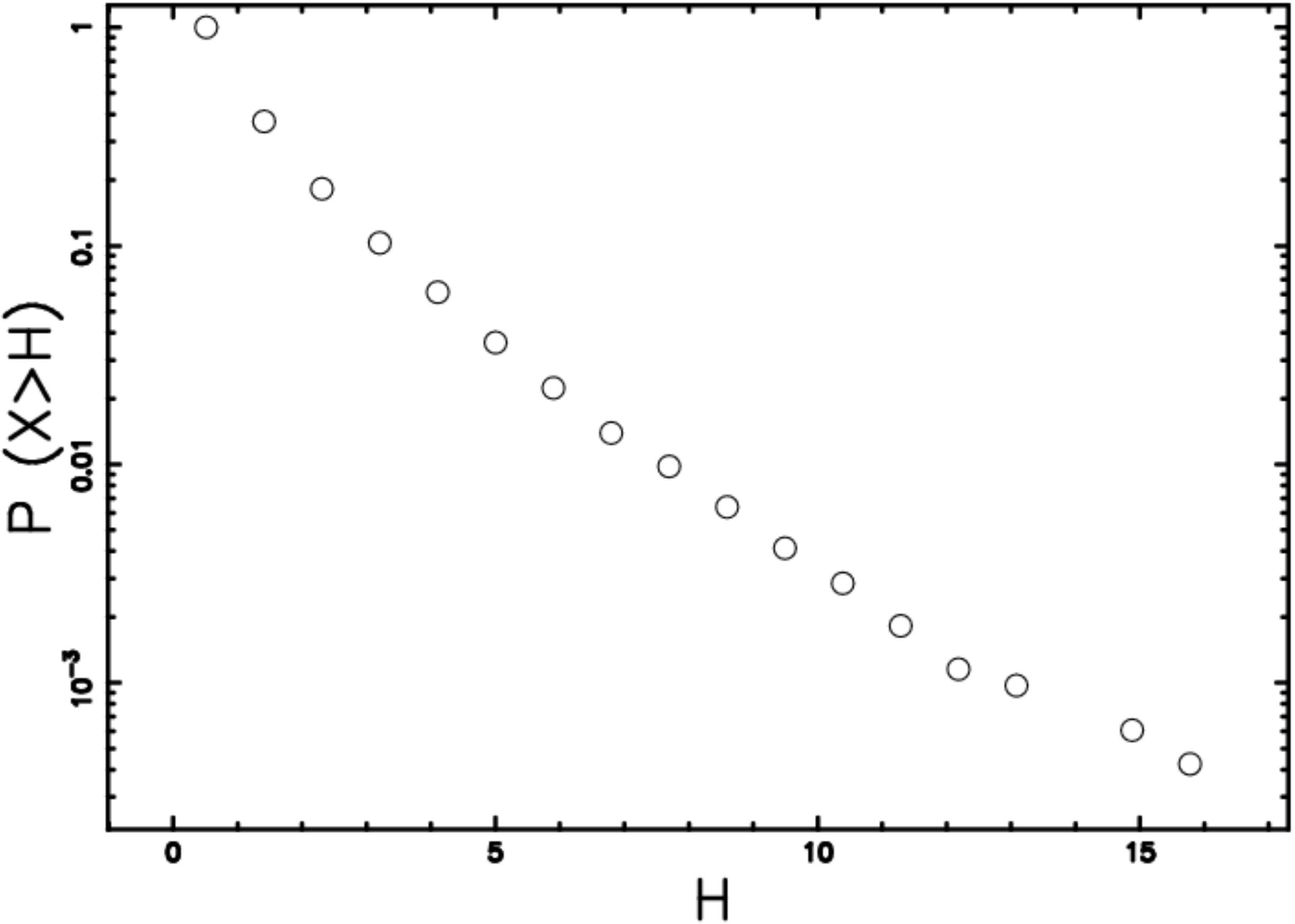}
\end {center}
\caption
{
Survival function of   of the ensemble
of  91 $\times$ 181  grid  points  as  
function of $H$ at   $D=56.9 Mpc$.
The median  is at $H$= 0.6 ,
in an homogeneous  sample  it should be at
$H$= 1.
}
          \label{survival}%
    \end{figure}

Figure  \ref{rho}  
which  shows  the  linear density 
contours of galaxies and  Figure \ref{comparison} 
which shows  the  
linear contours of  H are similar 
because the numbers of simulated  
galaxies in the selected cube   of 
the random sample
fluctuates  less   in respect to  the sample  of galaxies.

\section{Luminosity  effects}
\label{luminosity}

This section reports   the luminosity  function for galaxies,
the Malmquist bias which allows to build
 a complete sample
of galaxies,
and the numerical determination of the spatial dimension
of the LSC.

\subsection{Luminosity Function}

A model for the luminosity of galaxies
is  the Schechter function
\begin{equation}
\Phi (L) dL  = (\frac {\Phi^*}{L^*}) (\frac {L}{L^*})^{\alpha}
\exp \bigl ( {-  \frac {L}{L^*}} \bigr ) dL \quad ,
\label{equation_schechter}
\end {equation}
where $\alpha$ sets the slope 
for low values of $L$, $L^*$ is the
characteristic luminosity, and $\Phi^*$ is a normalization.
This
function  was suggested  by \cite{schechter} and
the distribution in absolute magnitude is
\begin{eqnarray}
\Phi (M)dM=&&(0.4  ln 10) \Phi^* 10^{0.4(\alpha +1 ) (M^*-M)}
\nonumber\\
&& \times \exp \bigl ({- 10^{0.4(M^*-M)}} \bigr)  dM \quad ,
\label{equation_schechter_M}
\end {eqnarray}
where $M^*$ is the characteristic magnitude
as derived from the
data.
The parameters  of the Schechter function
concerning the 2MRS
can be found in \cite{Cole2001}
and
are reported in   Table~\ref{parameters}.
\begin{table}
 \caption[]
{
The parameters of the Schechter function
and bolometric magnitude
for the 2MRS   in the $K_s-band$.
}
 \label{parameters}
 \[
 \begin{array}{lc}
 \hline
 \hline
 \noalign{\smallskip}
parameter            & 2MRS                            \\ \noalign{\smallskip}
M^* - 5\log_{10}h ~ [mags] &  ( -23.44 \pm 0.03) \\ \noalign{\smallskip}
\alpha                     &   -0.96  \pm 0.05                 \\ \noalign{\smallskip}
\Phi^* ~[h^3~Mpc^{-3}]     &   ((1.08   \pm 0.16)10^{-2})    \\ \noalign{\smallskip}
M^\odot_{K_S}              &   3.39                            \\ \noalign{\smallskip}
h                          &   0.7                              \\ \noalign{\smallskip}
 \hline
 \hline
 \end{array}
 \]
 \end {table}

\begin{figure}
\begin{center}
\includegraphics[width=6cm]{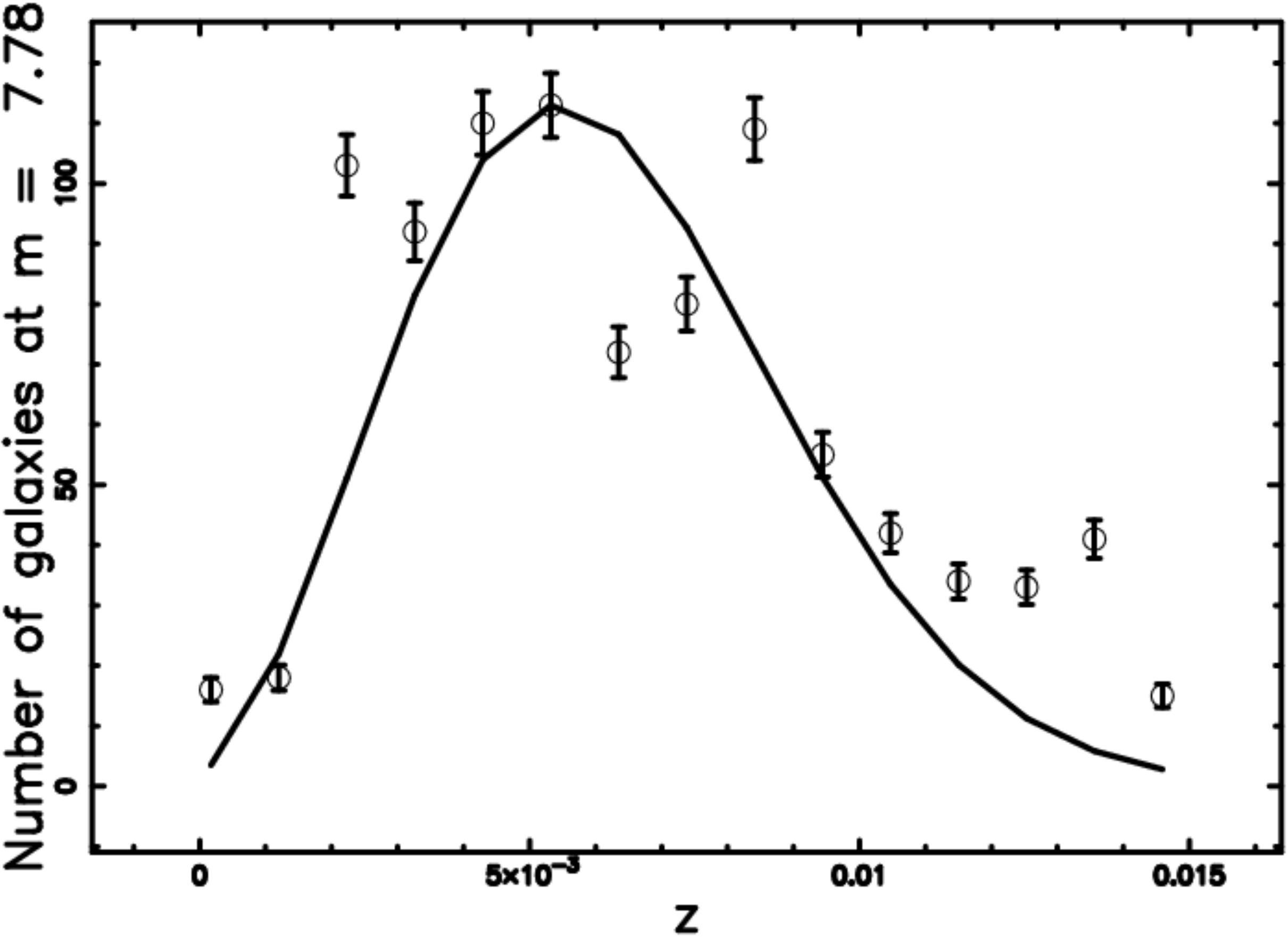}
\end {center}
\caption{
The galaxies  of the 2MRS with
$ 7.23  \leq  m  \leq 8.92 $  or
$  4881116  \frac {L_{\sun}}{Mpc^2} \leq
f \leq  23044324 \frac {L_{\sun}}{Mpc^2}$
are  organized in frequencies versus
heliocentric  redshift,
(empty circles);
the error bar is given by the square root of the frequency.
The maximum frequency of observed galaxies is
at  $z=0.0058$.
The full line is the theoretical curve  generated by (\ref{nfunctionz}), 
the Schechter  function of luminosity,
with  parameters
as in Table~\ref{parameters}.
In this plot $\mathcal{M_{\sun}}$ = 3.39  ,$h$ = 0.7
,  $\chi^2=132.93$ and the number of bins 15.
}
          \label{maximum_flux}%
    \end{figure}

The number of galaxies at a given flux  $f$
as  a function of the redshift  $z$,
see Formula~(1.104) in~\cite{pad},
 is
\begin{equation}
\frac{dN}{d\Omega dz df} =  4 \pi
\bigl ( \frac {c}{H_0} \bigr )^5    z^4 \Phi (\frac{z^2}{z_{crit}^2})
\label{nfunctionz}
\quad,
\end {equation}
where $d\Omega$, $dz$, and  $ df $
are the differentials of
the solid angle, the red-shift, and the flux, respectively,
and
\begin{equation}
 z_{crit}^2 = \frac {H_0^2  L^* } {4 \pi f c^2}
\quad,
\end{equation}
where $c$ is the velocity of light.
The number of galaxies at a given flux
has a maximum  at  $z=z_{max}$,
where
\begin{equation}
 z_{max} = z_{crit}  \sqrt {\alpha +2 }
\quad .
\label{posmaximum} 
\end{equation}
Figure~\ref{maximum_flux}
reports the number of  observed  galaxies
in  the 2MRS  catalog at a given
apparent magnitude  and
the theoretical curve  as represented by
Equation~(\ref{nfunctionz}).
The merit function $\chi^2$
can be computed as
\begin{equation}
\chi^2 =
\sum_{j=1}^n ( \frac {n_{theo}(z) -
                      n_{astr}(z) } {\sigma_{n_{astr}(z)}})^2
\quad ,
\label{chisquare}
\end{equation}
where   $n$ is number of data , the two indexes $theo$ and $astr$
stand for theoretical and astronomical, respectively and
 ${\sigma_{n_{astr}(z)}}^2$  is the variance of
 the astronomical number of data; the
obtained value is  reported in the caption of Figure
\ref{maximum_flux}.

On analyzing Figure~\ref{maximum_flux} we 
suggest  that  the disagreement between  theory and
observations  is due to the fact  that 
the number of galaxies does not increases 
exactly  with $R^3$  as  inserted  
in the theory but 
with  $R^{1.87}$ , see equation 
(\ref{numerodimensione}),
as  given by  the cellular structure.

\label{secflux}

\subsection{Malmquist  Bias}

\begin{figure*}
\begin{center}
\includegraphics[width=6cm]{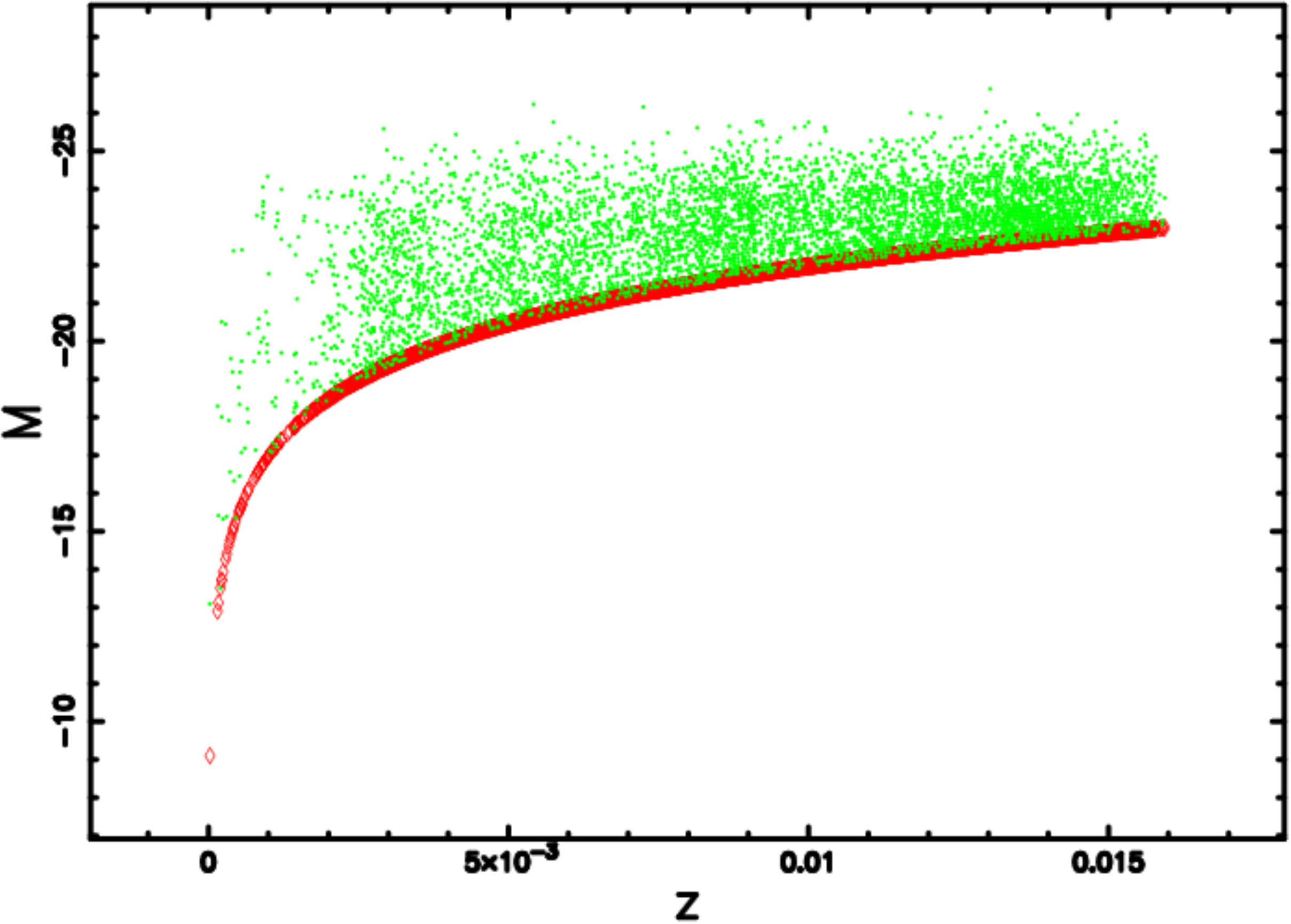}
\end {center}
\caption{
The absolute magnitude $M$ of
5512 galaxies belonging to the 2MRS
when $\mathcal{M_{\sun}}$ = 3.39 and
$H_{0}=70 \mathrm{\ km\ s}^{-1}\mathrm{\ Mpc}^{-1}$
(green points).
The upper theoretical curve as represented by
Equation~(\ref{absolutel}) is reported as the
red thick line when $m_L$=11.19.
}
 \label{bias}%
 \end{figure*}

The
 Malmquist bias,
see \cite{Malmquist_1920,Malmquist_1922},
 was originally applied
to the stars and later on 
 to the galaxies by \cite{Behr1951}.
The observable absolute magnitude as a function of the
limiting apparent magnitude, $m_L$, is
\begin{equation}
M_L =
m_{{L}}-5\,{\it \log_{10}} \left( {\frac {{\it c}\,z}{H_{{0}}}}
 \right) -25
\quad .
\label{absolutel}
\end{equation}
The previous formula predicts, from a theoretical
point of view, an upper limit on the maximum absolute  magnitude which  can be observed in a
catalog of galaxies characterized by a given limiting
magnitude and Figure~\ref{bias} reports such a curve
and the galaxies of the
2MRS.

The interval covered by the
LF of galaxies,
$\Delta M$,
is defined by
\begin{equation}
\Delta M = M_{max} - M_{min}
\quad,
\end{equation}
where $M_{max}$ and $M_{min}$ are the
maximum and minimum
absolute
magnitude of the LF for the considered catalog.
The real observable interval in absolute magnitude,
$\Delta M_L$,
 is
\begin{equation}
\Delta M_L = M_{L} - M_{min}
\quad .
\end{equation}
We can therefore introduce the range
of observable absolute maximum magnitude
expressed in percent,
$ \epsilon_s(z) $,
as
\begin{equation}
\epsilon_s(z) = \frac { \Delta M_L } {\Delta M } \times 100\%
\quad .
\label{range}
\end{equation}
This is a number which  represents
the completeness
of the sample
and, given the fact that the limiting magnitude of the 2MRS is
$m_L$=11.19,
it is possible to conclude that the 2MRS  is complete
for $z\leq0.00016$~.
This efficiency expressed as a
percentage
can be considered a version  of the Malmquist bias.

The previous treatment
does not take into account
the
$K$-corrections between observed ($Q$) and desired
($R$) bandpasses,
$K_{QR}(z)$ , see equation (1)  \cite{Blanton2007}.
The
$K$-correction
depends  from the chosen catalog and bandpass .
In our case   we  are interested in band $K_s$
in the 2MASS and  the following
approximate  relationship  can be drawn from
Figure 14  in  \cite{Blanton2007}:
\begin{equation}
K_K(z) \approx  1.2  z
\quad  .
\label{kz}
\end{equation}
As an example  at $z \approx$ 0.016 , which is the maximum
processed value of redshift  of the LSC,
 the $K$-corrections are
0.02 mag. This means that the $K$-corrections are negligible when
applied to the LSC , which is a subset of the 2MASS.

\subsection{The Radial Distribution of Galaxies}
\label{secbeta}
\begin{figure}
\begin{center}
\includegraphics[width=6cm]{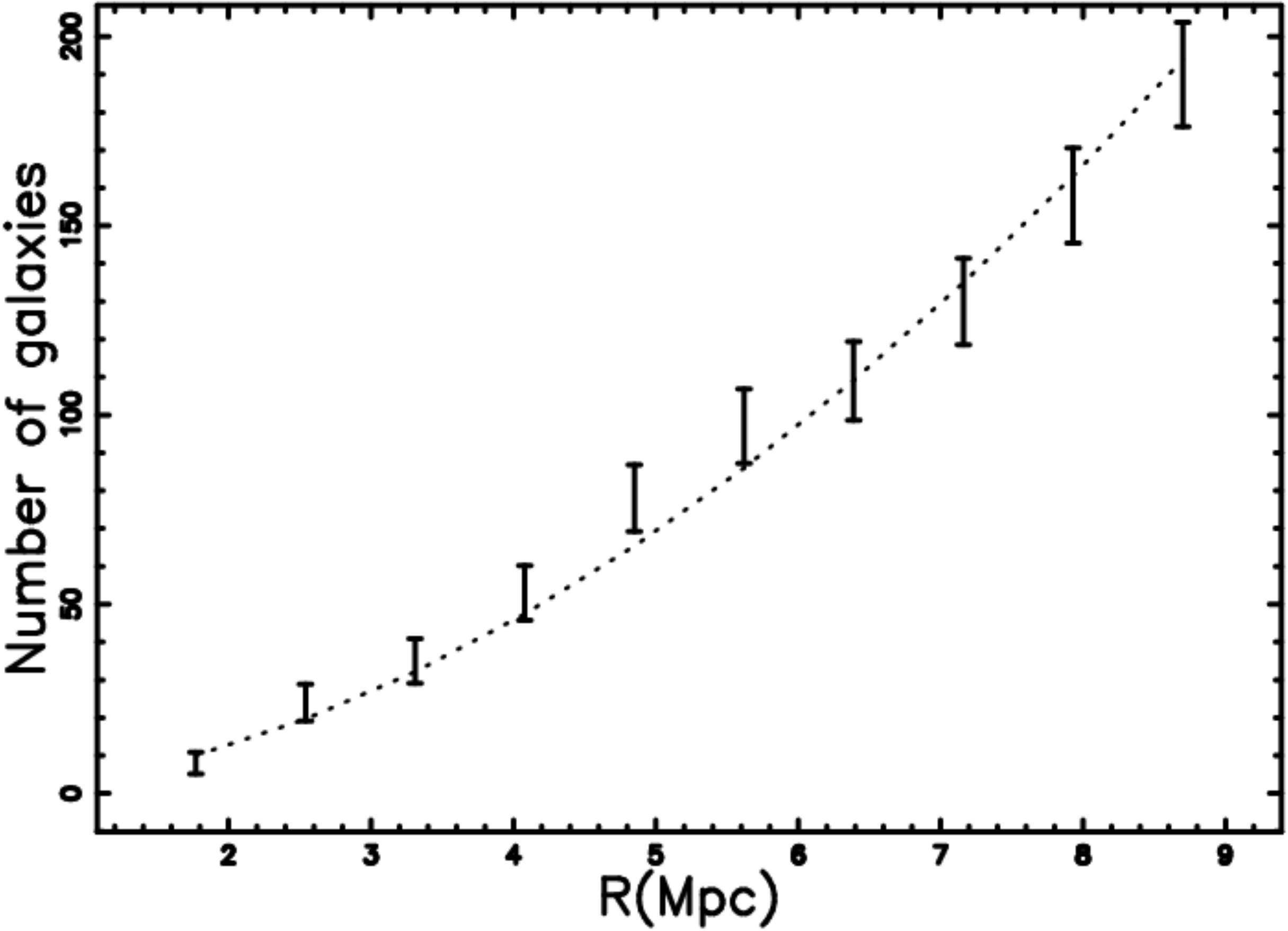}
\end {center}
\caption
{
The  number of  galaxies  as  function of the distance  
in Mpc for  the  2MRS catalog (full line)
and  observed  number of galaxies 
with
vertical error bars.
The  error  is computed as the square root of the number 
of galaxies in the considered bin.
}
          \label{dimension}%
    \end{figure}

In a 3D random distribution,   the number of galaxies grows
$\propto  R^3$ where $R$  is the distance  from
the center (our galaxy) in Mpc.
In a framework of galaxies situated on the first face  of
a PVT  their  number
should grow   $\propto  R^2$
assuming   that $R$ denotes the distance
from the center of the face.
The counting of the nearest galaxies can be done in light
of the
Malmquist bias and the first face of a PVT.
According to (\ref{firstface}),
the  galaxies on  the first face
will terminate
at  $D_F \approx 8.75 \,Mpc$
which corresponds to $z=0.00204$.
The Malmquist line as represented by
(\ref{absolutel})
suggests that the sample of galaxies
in the interval $0 < z < 0.00204$  and 
characterized $ M < -18.51 $    is complete.
According  to equation (\ref{rpower})  
the spatial behavior is
\begin{eqnarray}
\label{numerodimensione}
N(R) = (3.52\pm 0.49)  R^{(1.87\pm0.09)}  \\
\quad when
\quad
0 < z < 0.00204
\quad
and
\quad
M < -18.51
\quad ,\nonumber
\end{eqnarray}
and  Figure \ref{dimension} reports  such a relationship.

\begin{figure}
\begin{center}
\includegraphics[width=6cm]{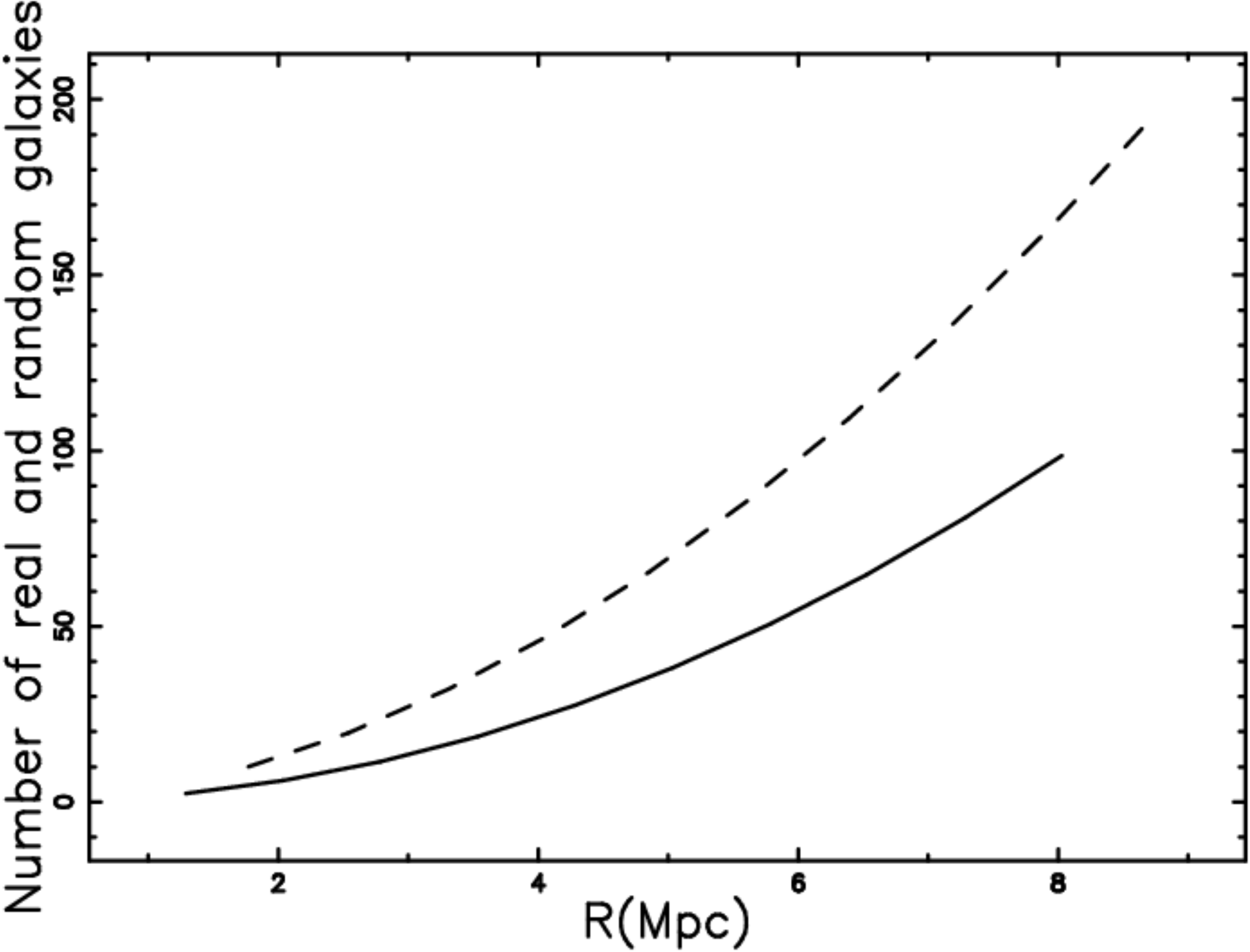}
\end {center}
\caption
{
The  number of  real galaxies  as  function of the distance  
in Mpc for  the  2MRS catalog (full line)
and  number of random galaxies generated on a plane
(dashed line).
}
          \label{dimensionran}%
    \end{figure}

This  last  result can be tested on a random 2D
sample in which the galaxies are generated
on a plane having the same area of the astronomical
analyzed region.
The  result  of the simulation for the random galaxies gives
\begin{eqnarray}
N(R) = (1.43 \pm 0.03)  R^{(2.03\pm0.082)}  \\
\quad when
\quad
0 < z < 0.00204
\quad
.
\nonumber
\end{eqnarray}
A comparison of the two curves of
growth ( random galaxies and observed galaxies)
is reported in Figure \ref{dimensionran}.

Others  approaches   use the  generalized 
fractal dimension  estimator to  
(i) analyze the scaling 
properties of the 
clusters of galaxies, 
see \cite{Borgani1993},
(ii) determine the scale of homogeneity in the
Universe as  derived from various galaxy redshift surveys,
see \cite{Bagla2008},
(iii)  study   the transition to homogeneity,
defined as the scale above which
the fractal dimension  of the underlying 
galaxies distribution is equal to
the ambient dimension  of the space in 
which the galaxies  are distributed, see 
\cite{Yadav2010}.

\section{Conclusions}
\label{conclusions}

In  Section \ref{catalogue} we  learned
that the concept of homogeneity does not
correspond to the  observations  of LSC   
because the  63~$\%$ of 
$91\times181$
grid points  processed 
at a  distance of $\approx$ 59 Mpc  
have the indicator of homogeneity lower than 1. 
The concept
 of a non-isotropic universe is a consequence
of the previous statement, in other
words, the  appearance of the universe depends
on the chosen observer/galaxy.
These facts, based on the theoretical and
observational bases,
allow  us  to assert
a local new  version of the Cosmological Principle
``The galaxies, as well our galaxy, are situated
on the faces of a Voronoi 
polyhedron. The spatial distribution
of galaxies will follow approximately the
geometrical rules as well the standard photometric
rules.''
A first set of  consequences  can be drawn
\begin{enumerate}
\item  The number of galaxies
in the first face of a PVT should grown as
$N(R)  \propto R^2$.
An analysis of the number of galaxies in the first
8.75 Mpc of the LSC gives
$N(R)  \propto R^{1.87}$.
\item A  new definition
of  the density of galaxies
on the faces of a PVT, $n_f$, which in the framework
of thick faces is
\begin{equation}
n_f \approx n \frac{1}{3 \rho}
\quad ,
\end{equation}
where $\rho =s / \bar{R}$  is a dimensional parameter
lower than 1, $s$ is the thickness of the faces, and
$n$ is the usual density  of galaxies.
\item
The so called ``filaments of galaxies'' are here
classified as 2D sections of a 3D PVT which
are made by irregular polygons.
\item
The reformulation here presented is in agreement with
the anisotropies in the short and intermediate
gamma ray bursts which are explained
by a breakdown of the
Cosmological Principle, see
\cite{Meszaros2009a},\cite{Meszaros2009b}.
\item 
The theoretical  dimension which regulates 
the number of galaxies as function of the distance 
is  different when the center of the selected 
sphere belongs  to a thick face or a thick edge.
\end{enumerate}
Different approaches   which  analyze  the
SDSS  
seem to suggest 
the validity of cosmological principle
and they  are:
(i)  the  test the homogeneity of the universe at 
     z $\approx$ 0.3 
     with the luminous red galaxy (LRG) spectroscopic 
     sample of the SDSS, see \cite{Hogg2005},
(ii) 
a multifractal
analysis of the galaxy distribution in a volume-limited 
sub-sample from the (SDSS)
which  finds that the galaxy distribution
becomes homogeneous at a length-scale between 60/h Mpc 
and 70/h Mpc, see \cite{Sarkar2009}.
We remember that 70/h Mpc as a linear dimension
comprises $\approx$ 2 PVT cells.
The analysis here presented is  limited to  galaxies 
within 8 Mpc
in order to  measure the  spatial dimension.
Tests carried out by other authors also do not find homogeneity
within 8 Mpc. 
Rather, the  transition to homogeneity is found to
occur at a much larger length-scale. 
So there does not appear to
be any inconsistency with the findings of our  paper and the
currently accepted picture of the transition to homogeneity.
In the framework of the cellular universe 
the transition  from  non homogeneous to homogeneous
regime  is   reached   when  the  cube in which 
the density of galaxies is  evaluated comprises many cells.


\begin{thebibliography}{36}
\expandafter\ifx\csname natexlab\endcsname\relax\def\natexlab#1{#1}\fi

\bibitem[{{Bagla} {et~al.}(2008){Bagla}, {Yadav}, \& {Seshadri}}]{Bagla2008}
{Bagla}, J.~S., {Yadav}, J., \& {Seshadri}, T.~R. 2008, \mnras, 390, 829

\bibitem[{{Behr}(1951)}]{Behr1951}
{Behr}, A. 1951, Astronomische Nachrichten, 279, 97

\bibitem[{{Blanton} \& {Roweis}(2007)}]{Blanton2007}
{Blanton}, M.~R., \& {Roweis}, S. 2007, \aj, 133, 734

\bibitem[{{Borgani} {et~al.}(1993){Borgani}, {Murante}, {Provenzale}, \&
  {Valdarnini}}]{Borgani1993}
{Borgani}, S., {Murante}, G., {Provenzale}, A., \& {Valdarnini}, R. 1993, \pre,
  47, 3879

\bibitem[{{Brynjolfsson}(2009)}]{Brynjolfsson2009}
{Brynjolfsson}, A. 2009, in Astronomical Society of the Pacific Conference
  Series, Vol. 413, Astronomical Society of the Pacific Conference Series, ed.
  {F.~Potter}, 169--189

\bibitem[{{Cole} {et~al.}(2001){Cole}, {Norberg}, {Baugh}, {Frenk},
  {Bland-Hawthorn}, {Bridges}, {Cannon}, {Colless}, {Collins}, {Couch},
  {Cross}, {Dalton}, {De Propris}, {Driver}, {Efstathiou}, {Ellis},
  {Glazebrook}, {Jackson}, {Lahav}, {Lewis}, {Lumsden}, {Maddox}, {Madgwick},
  {Peacock}, {Peterson}, {Sutherland}, \& {Taylor}}]{Cole2001}
{Cole}, S., {Norberg}, P., {Baugh}, C.~M., {Frenk}, C.~S., {Bland-Hawthorn},
  J., {Bridges}, T., {Cannon}, R., {Colless}, M., {Collins}, C., {Couch}, W.,
  {Cross}, N., {Dalton}, G., {De Propris}, R., {Driver}, S.~P., {Efstathiou},
  G., {Ellis}, R.~S., {Glazebrook}, K., {Jackson}, C., {Lahav}, O., {Lewis},
  I., {Lumsden}, S., {Maddox}, S., {Madgwick}, D., {Peacock}, J.~A.,
  {Peterson}, B.~A., {Sutherland}, W., \& {Taylor}, K. 2001, \mnras, 326, 255

\bibitem[{{Colless} {et~al.}(2001){Colless}, {Dalton}, {Maddox}, \& {et
  al.}}]{Colless2001}
{Colless}, M., {Dalton}, G., {Maddox}, S., \& {et al.} 2001, \mnras, 328, 1039

\bibitem[{{Crook} {et~al.}(2007){Crook}, {Huchra}, {Martimbeau}, {Masters},
  {Jarrett}, \& {Macri}}]{Huchra2007}
{Crook}, A.~C., {Huchra}, J.~P., {Martimbeau}, N., {Masters}, K.~L., {Jarrett},
  T., \& {Macri}, L.~M. 2007, \apj, 655, 790

\bibitem[{{Geller} \& {Huchra}(1989)}]{geller}
{Geller}, M.~J., \& {Huchra}, J.~P. 1989, Science, 246, 897

\bibitem[{{Hogg} {et~al.}(2005){Hogg}, {Eisenstein}, {Blanton}, {Bahcall},
  {Brinkmann}, {Gunn}, \& {Schneider}}]{Hogg2005}
{Hogg}, D.~W., {Eisenstein}, D.~J., {Blanton}, M.~R., {Bahcall}, N.~A.,
  {Brinkmann}, J., {Gunn}, J.~E., \& {Schneider}, D.~P. 2005, \apj, 624, 54

\bibitem[{{Hubble}(1929)}]{Hubble1929}
{Hubble}, E. 1929, Proceedings of the National Academy of Science, 15, 168

\bibitem[{{Huchra} {et~al.}(2012){Huchra}, {Macri}, {Masters}, \&
  {et~al.}}]{Huchra2012}
{Huchra}, J.~P., {Macri}, L.~M., {Masters}, K.~L., \& {et~al.} 2012, \apjs,
  199, 26

\bibitem[{{Jarrett}(2004)}]{Jarrett2004}
{Jarrett}, T. 2004, \pasa, 21, 396

\bibitem[{{Jones} {et~al.}(2004){Jones}, {Saunders}, {Colless}, {Read}, \&
  {Parker}}]{Jones2004}
{Jones}, D.~H., {Saunders}, W., {Colless}, M., {Read}, M.~A., \& {Parker},
  Q.~A.~e. 2004, \mnras, 355, 747

\bibitem[{{Keel}(2007)}]{Keel2007}
{Keel}, W.~C. 2007, {The road to galaxy formation} (Berlin: Springer)

\bibitem[{{Keselman} {et~al.}(2010){Keselman}, {Nusser}, \&
  {Peebles}}]{Peebles2010}
{Keselman}, J.~A., {Nusser}, A., \& {Peebles}, P.~J.~E. 2010, \prd, 81, 063521

\bibitem[{{Liddle} \& {Loveday}(2009)}]{Oxford_Cosmology_2009}
{Liddle}, A., \& {Loveday}, J. 2009, {The Oxford Companion to Cosmology}
  (Oxford: Oxford University Press)

\bibitem[{{Malmquist }(1920)}]{Malmquist_1920}
{Malmquist }, K. 1920, Lund Medd. Ser.~II, 22, 1

\bibitem[{{Malmquist }(1922)}]{Malmquist_1922}
---. 1922, Lund Medd. Ser.~I, 100, 1

\bibitem[{{Meszaros} {et~al.}(2009){Meszaros}, {Balazs}, {Bagoly}, \&
  {Veres}}]{Meszaros2009a}
{Meszaros}, A., {Balazs}, L.~G., {Bagoly}, Z., \& {Veres}, P. 2009, in American
  Institute of Physics Conference Series, Vol. 1133, American Institute of
  Physics Conference Series, ed. {C.~Meegan, C.~Kouveliotou, \& N.~Gehrels},
  483--485

\bibitem[{{M{\'e}sz{\'a}ros} {et~al.}(2009){M{\'e}sz{\'a}ros}, {Bal{\'a}zs},
  {Bagoly}, \& {Veres}}]{Meszaros2009b}
{M{\'e}sz{\'a}ros}, A., {Bal{\'a}zs}, L.~G., {Bagoly}, Z., \& {Veres}, P. 2009,
  Baltic Astronomy, 18, 293

\bibitem[{{Okabe} {et~al.}(1992){Okabe}, {Boots}, \& {Sugihara}}]{okabe}
{Okabe}, A., {Boots}, B., \& {Sugihara}, K. 1992, {Spatial tessellations.
  Concepts and Applications of Voronoi diagrams} ({Chichester, New York}:
  {Wiley})

\bibitem[{{Padmanabhan}(1996)}]{pad}
{Padmanabhan}, T. 1996, Cosmology and Astrophysics through Problems (Cambridge:
  Cambridge University Press)

\bibitem[{{Pan} {et~al.}(2012){Pan}, {Vogeley}, {Hoyle}, {Choi}, \&
  {Park}}]{Vogeley2012}
{Pan}, D.~C., {Vogeley}, M.~S., {Hoyle}, F., {Choi}, Y.-Y., \& {Park}, C. 2012,
  \mnras, 421, 926

\bibitem[{{Press} {et~al.}(1992){Press}, {Teukolsky}, {Vetterling}, \&
  {Flannery}}]{press}
{Press}, W.~H., {Teukolsky}, S.~A., {Vetterling}, W.~T., \& {Flannery}, B.~P.
  1992, {Numerical Recipes in FORTRAN. The Art of Scientific Computing}
  (Cambridge: Cambridge University Press)

\bibitem[{{Ryden}(2003)}]{Ryden2003}
{Ryden}, B. 2003, {Introduction to Cosmology} (San Francisco, CA, USA: Addison
  Wesley)

\bibitem[{{Sarkar} {et~al.}(2009){Sarkar}, {Yadav}, {Pandey}, \&
  {Bharadwaj}}]{Sarkar2009}
{Sarkar}, P., {Yadav}, J., {Pandey}, B., \& {Bharadwaj}, S. 2009, \mnras, 399,
  L128

\bibitem[{{Schechter}(1976)}]{schechter}
{Schechter}, P. 1976, \apj, 203, 297

\bibitem[{{Schwarz}(2009)}]{Schwarz2009}
{Schwarz}, D.~J. 2009, eprint arXiv:0905.0384

\bibitem[{{van de Weygaert} \& {Icke}(1989)}]{Weygaert1989}
{van de Weygaert}, R., \& {Icke}, V. 1989, \aap, 213, 1

\bibitem[{{Yadav} {et~al.}(2010){Yadav}, {Bagla}, \& {Khandai}}]{Yadav2010}
{Yadav}, J.~K., {Bagla}, J.~S., \& {Khandai}, N. 2010, \mnras, 405, 2009

\bibitem[{{Zaninetti}(1991)}]{zaninettig}
{Zaninetti}, L. 1991, \aap, 246, 291

\bibitem[{{Zaninetti}(1995)}]{zaninetti95}
---. 1995, \aaps, 109, 71

\bibitem[{{Zaninetti}(2006)}]{Zaninetti2006}
---. 2006, \cjaa, 6, 387

\bibitem[{{Zaninetti}(2009)}]{Zaninetti2009c}
---. 2009, \pla, 373, 3223

\bibitem[{{Zaninetti}(2010)}]{Zaninetti2010a}
---. 2010, Revista Mexicana de Astronomia y Astrofisica, 46, 115

\end{thebibliography}

\end{document}